\documentclass[article,reprint,superscriptaddress,longbibliography]{revtex4-1}
\usepackage{graphicx}
\usepackage{dcolumn}
\usepackage{bm}
\usepackage{hyperref} 
\usepackage{breakurl}
\usepackage{multirow}
\usepackage{enumitem}
\usepackage{color}
\usepackage[pagewise]{lineno}
\setcitestyle{super} 

\begin{document}

\title{Bulk Ising superconductivity in an intercalated TaSe$_{2}$ bilayer structure}

\author{Yupeng Li}
      \thanks{Equal contributions}
      \email{physyupengli@hznu.edu.cn}
      \affiliation{Hangzhou Key Laboratory of Quantum Matter, School of Physics, Hangzhou Normal University, Hangzhou 311121, China}
      \affiliation{Beijing National Laboratory for Condensed Matter Physics, Institute of Physics, Chinese Academy of Sciences, Beijing 100190, China}
      \affiliation{School of Physics, Zhejiang University, Hangzhou 310058, China}

\author{Zhaopeng Guo}
      \thanks{Equal contributions}
      \affiliation{Hangzhou Key Laboratory of Quantum Matter, School of Physics, Hangzhou Normal University, Hangzhou 311121, China}

\author{Lihong Hu}
      \affiliation{Beijing National Laboratory for Condensed Matter Physics, Institute of Physics, Chinese Academy of Sciences, Beijing 100190, China}
      \affiliation{School of Physical Sciences, University of Chinese Academy of Sciences, Beijing 100049, China}

\author{Guoan Li}
      \affiliation{Beijing National Laboratory for Condensed Matter Physics, Institute of Physics, Chinese Academy of Sciences, Beijing 100190, China}

\author{Siqi Wu}
      \affiliation{School of Physics, Zhejiang University, Hangzhou 310058, China}

\author{Xinyi Zheng}
      \affiliation{Beijing National Laboratory for Condensed Matter Physics, Institute of Physics, Chinese Academy of Sciences, Beijing 100190, China}
      \affiliation{School of Physical Sciences, University of Chinese Academy of Sciences, Beijing 100049, China}

\author{Xiao Deng}
      \affiliation{Beijing National Laboratory for Condensed Matter Physics, Institute of Physics, Chinese Academy of Sciences, Beijing 100190, China}
      \affiliation{School of Physical Sciences, University of Chinese Academy of Sciences, Beijing 100049, China}

\author{Zhiyuan Zhang}
      \affiliation{Beijing National Laboratory for Condensed Matter Physics, Institute of Physics, Chinese Academy of Sciences, Beijing 100190, China}
      \affiliation{School of Physical Sciences, University of Chinese Academy of Sciences, Beijing 100049, China}

\author{Anqi Wang}
      \affiliation{Beijing National Laboratory for Condensed Matter Physics, Institute of Physics, Chinese Academy of Sciences, Beijing 100190, China}

\author{Xingchen Guo}
      \affiliation{Beijing National Laboratory for Condensed Matter Physics, Institute of Physics, Chinese Academy of Sciences, Beijing 100190, China}
      \affiliation{School of Physical Sciences, University of Chinese Academy of Sciences, Beijing 100049, China}

\author{Ziwei Dou}
      \affiliation{Beijing National Laboratory for Condensed Matter Physics, Institute of Physics, Chinese Academy of Sciences, Beijing 100190, China}

\author{Peiling Li}
      \affiliation{Beijing National Laboratory for Condensed Matter Physics, Institute of Physics, Chinese Academy of Sciences, Beijing 100190, China}

\author{Yuke Li}
      \affiliation{Hangzhou Key Laboratory of Quantum Matter, School of Physics, Hangzhou Normal University, Hangzhou 311121, China}

\author{Fanming Qu}
      \affiliation{Beijing National Laboratory for Condensed Matter Physics, Institute of Physics, Chinese Academy of Sciences, Beijing 100190, China}
      \affiliation{School of Physical Sciences, University of Chinese Academy of Sciences, Beijing 100049, China}

\author{Guangtong Liu}
      \affiliation{Beijing National Laboratory for Condensed Matter Physics, Institute of Physics, Chinese Academy of Sciences, Beijing 100190, China}
      \affiliation{Songshan Lake Materials Laboratory, Dongguan 523808, China}

\author{Jin-Ke Bao}
      \email{jinke$_$bao@hznu.edu.cn}
      \affiliation{Hangzhou Key Laboratory of Quantum Matter, School of Physics, Hangzhou Normal University, Hangzhou 311121, China}

\author{Guang-Han Cao}
      \affiliation{School of Physics, Zhejiang University, Hangzhou 310058, China}
      \affiliation{State Key Laboratory of Silicon and Advanced Semiconductor Materials, Zhejiang University, Hangzhou 310058, China}
      \affiliation{Collaborative Innovation Centre of Advanced Microstructures, Nanjing University, Nanjing 210093, China}

\author{Li Lu}
      \affiliation{Beijing National Laboratory for Condensed Matter Physics, Institute of Physics, Chinese Academy of Sciences, Beijing 100190, China}
      \affiliation{School of Physical Sciences, University of Chinese Academy of Sciences, Beijing 100049, China}
      \affiliation{Songshan Lake Materials Laboratory, Dongguan 523808, China}

\author{Jie Shen}
      \email{shenjie@iphy.ac.cn}
      \affiliation{Beijing National Laboratory for Condensed Matter Physics, Institute of Physics, Chinese Academy of Sciences, Beijing 100190, China}
      \affiliation{Songshan Lake Materials Laboratory, Dongguan 523808, China}

\author{Zhu-An Xu}
      \email{zhuan@zju.edu.cn}
      \affiliation{School of Physics, Zhejiang University, Hangzhou 310058, China}
      \affiliation{State Key Laboratory of Silicon and Advanced Semiconductor Materials, Zhejiang University, Hangzhou 310058, China}
      \affiliation{Collaborative Innovation Centre of Advanced Microstructures, Nanjing University, Nanjing 210093, China}

\begin{abstract}
Ising spin-orbit coupling in bulk systems has drawn considerable interest for its ability to conveniently construct spin-orbit environments and enable exotic quantum phenomena. In this work, we synthesize intercalated 2Hb-TaSe$_2$ bilayers with noncentrosymmetric structure and, through multifaceted analysis, present multiple lines of evidence for the emergence of bulk Ising superconductivity. Resistivity measurements reveal anisotropic superconducting behavior, with a remarkably large in-plane upper critical field $B_{c2}^{\|}$ that exceeds the Pauli limit $B_{p}$. Band structure calculations further show band splitting accompanied by out-of-plane spin polarization. Collectively, these observations point to the presence of Ising superconductivity. Additional measurements of the thickness-dependent ratio $B_{c2}^{\|}$/$B_{p}$ and the superconducting diode effect not only further support the Ising superconducting nature of this material, but also reveal additional features of bulk Ising superconductivity evolving with thickness. Our findings provide valuable insights that may contribute to the search for bulk Ising superconductors.
\end{abstract}

\maketitle

\noindent\textbf{1 Introduction}
\vspace{3ex}

Since the discovery of out-of-plane Zeeman-type spin-orbit coupling (termed Ising SOC) in monolayer transition metal dichalcogenides (TMDs) \cite{MoS2IsingSC_Science15,MoS2IsingSC_NP16,NbSe22D_XiX_NP16}, this distinctive form of SOC and the associated Ising superconductivity (IS) have attracted considerable attention.
This interest is primarily driven by the exceptionally large in-plane upper critical field ($B_{c2}^{\|}$) arising from broken in-plane inversion symmetry, which enables significant phenomena such as orbital Fulde-Ferrell-Larkin-Ovchinnikov (FFLO) states \cite{FFLOinIsingSC_Nature23}, superconducting diode effects (SDE) \cite{NbSe2SCE_BauriedlL_NC2022}, topological superconductivity and Majorana fermions \cite{IsingSCMajorana_PRB2016,TSCinTMD_NC2018,TSCusinginIsingSC_NJP2020}.
However, Ising SOC has not been widely exploited as a modular structural motif or a tunable element for engineering complex quantum phenomena, in contrast to Rashba SOC, which can be readily induced or enhanced through dimensional (D) confinement to 2D \cite{2DSC_SatioY_NRM2017} or even 1D \cite{ZBPAlInAs_NP2012,Yplisummary_AQT19}, or via interface engineering such as surface deposition \cite{Y3Fe5O12FFJDE_JeonKR_NM2022}.
A primary limitation maybe stem from its reliance on ultrathin geometries (eg. monolayers), which are often experimentally challenging to integrate into heterostructures.
Furthermore, achieving high-quality and clean-limit superconductivity in certain materials sometimes remains difficult, complicating the disentanglement of intrinsic Ising SOC and spin-orbit scattering (SOS) in enhancing $B_{c2}^{\|}$. Additionally, structures containing monolayer IS unit do not uniformly exhibit IS \cite{TaSe2_Yokota_2000,InTaSe2_YpLi,InTaS2_YPLi_PRB2020,PbSeTaSe2_SST2018}. Collectively, these factors constrain the further experimental utilization of Ising SOC.

\begin{figure*}[!thb]
\begin{center}
\includegraphics[width=6.5in]{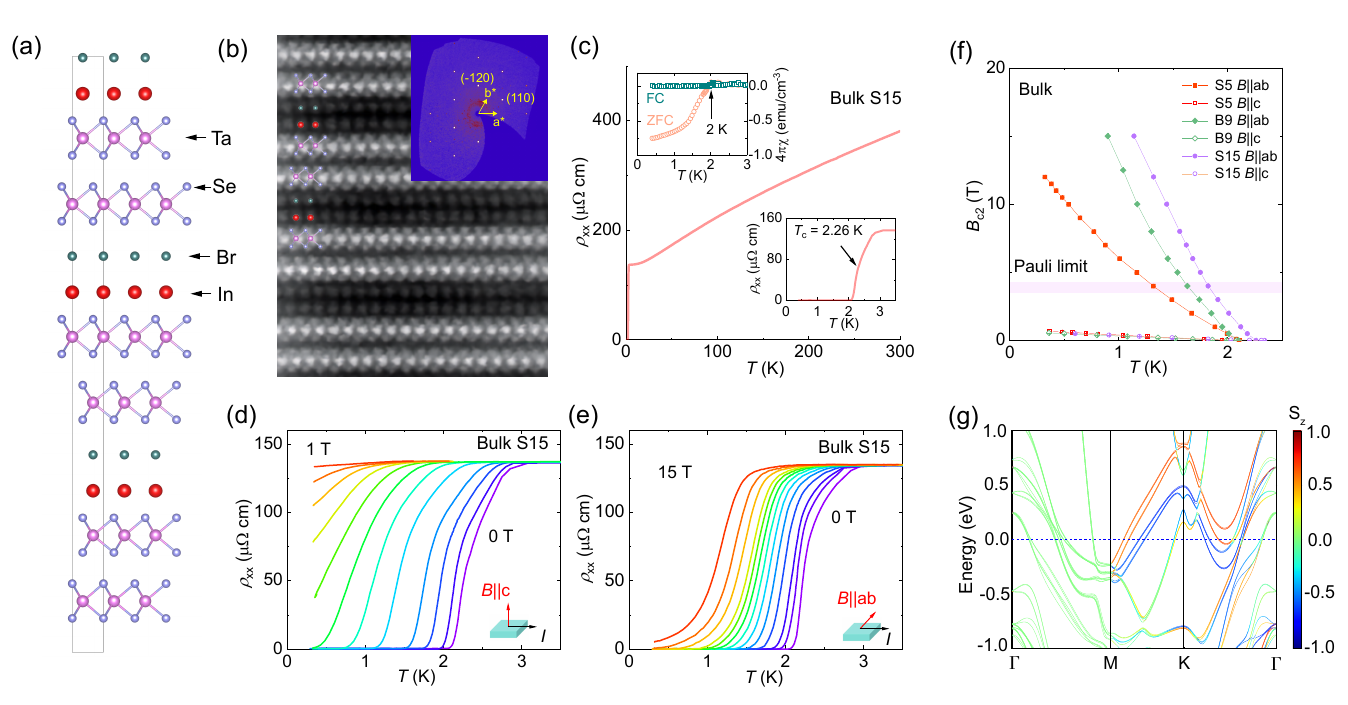}
\end{center}
\caption{\label{Fig1} Structure and superconductivity of bulk InBr(TaSe$_{2}$)$_{2}$. (a) Side view of the crystalline structure, highlighting the alternating 2Hb-TaSe$_{2}$ bilayers and InBr spacer layers.
(b) HRTEM image of InBr(TaSe$_{2}$)$_{2}$ taken along the [100] zone axis, with an overlaid atomic model. The inset shows the reconstructed image for ($hk0$) planes from SXRD data. (c) $R$-$T$ curve with a magnified view of the transition (lower right inset) and the corresponding magnetic susceptibility (upper left inset).
(d, e) Superconducting transitions under $B^{\perp}$ and $B^{\|}$, respectively.
(f) $T$-dependent $B_{c2}^{\perp}$ and $B_{c2}^{\|}$ for three bulk samples.
(g) Band structure of bulk InBr(TaSe$_2$)$_2$, with spin-up ($S_z > 0$) and spin-down ($S_z < 0$) components represented by red and blue colors, respectively.
 }
\end{figure*}

Therefore, investigating IS in bulk (also termed 3D IS) will facilitate the integration of Ising SOC with other intriguing physical phenomena. Several TMDs candidates with 3D IS have been reported through various methods, including intercalating large spacer layers - such as ionic liquids \cite{TaS2intercalation_science1970,TaS2interC2DSC_PRB1980,IntercalatedTaSe2_PRB1983,3DISingNbSe2Liquid_NP22} and other layered compounds \cite{NbS2Ba3NbS5_science20} - to increase the distance between 2D superconducting monolayer, as well as synthesizing intrinsic bulk materials \cite{4HbTaS2largeHc2_PRB2024,3DIsing4HbTaS2_NC2025,4HaNbSe2Ising_PRL2026,4HNbSe2Ising_PRL2025,3RTaSe2_AFM2025}.
A major challenge of identifying 3D IS lies in the fact that several candidates readily exhibit dirty-limit superconductivity, which may explain why such systems appear to be uncommon.

In this work, we examine the role of Ising SOC in enhancing the $B_{c2}^{\|}$ in InBr(TaSe$_{2}$)$_{2}$. Using a vapor transport method \cite{InTaSe2_YpLi}, we intercalate an InBr layer into superconducting 2Hb-TaSe$_{2}$ bilayers (Table SI; Fig. S1). This specific stacking of 2D Ising superconducting motifs is designed to effectively promote a large $B_{c2}^{\|}$, consistent with earlier reports of stacking-enhanced critical fields \cite{TaS2intercalation_science1970,TaS2interC2DSC_PRB1980,IntercalatedTaSe2_PRB1983}.
To overcome the ambiguity introduced by SOS \cite{SOScatterSC_PRB1975}, we employ multiple complementary experimental approaches to distinguish effect of Ising SOC, going beyond the mere observation of a large $B_{c2}^{\|}$ or classification within clean/dirty limits.

\vspace{3ex}
\noindent\textbf{2 Results and Discussion}

\noindent\textbf{2.1 Bulk properties}

InBr(TaSe$_{2}$)$_{2}$ single-crystal is synthesized with a rhombohedral structure, \emph{R3m} (space group no. 160), as shown in Fig. 1a.
This noncentrosymmetric structure is confirmed by high-resolution transmission electron microscopy (HRTEM, Fig. 1b) and single-crystal X-ray diffraction (SXRD, inset of Fig. 1b).
The InBr double layers consist of two monolayers of In and Br atoms as the dominant configuration, while also displaying a certain degree of randomness in the mixed atomic occupations. This disorder likely arises from variations in the actual atomic configurations of the double layer along the $c$-axis. This uncertainty is also underscored by the fact that even a structural model positing the complete absence of Br yields a similar refinement quality. Detailed refinement results are provided in Table SI.
For simplicity, the crystal structure depicted here is based on a model with fully ordered In and Br atoms within the InBr layer.
The unit cell has lattice parameters $a$ = $b$ = 3.451 \AA \ and $c$ = 57.353 \AA. The stoichiometric ratio In:Br:Ta:Se $\sim$ 1:1:2:4 is corroborated by energy-dispersive X-ray spectroscopy and inductively coupled plasma atomic emission spectrometry (Fig. S1).

Superconductivity of InBr(TaSe$_{2}$)$_{2}$ is systematically characterized.
Figure 1(c) reveals the superconducting transition temperature $T_{c}$ = 2.26 K,
determined by a 50\% drop in normal resistance [$R_{N}$, lower right inset of Fig. 1(c)].
This $T_{c}$ value is approximately consistent with that extracted in magnetic susceptibility measurements [upper left inset of Fig. 1(c)].
The evolution of $T_{c}$ under out-of-plane ($B {\perp} ab$) and in-plane
($B {\|} ab$ plane) magnetic fields is presented in Figs. 1(d) and 1(e), respectively.
A pronounced anisotropy between $B_{c2}^{\perp}$ and $B_{c2}^{\|}$ is clearly observed for three bulk samples in Fig. 1(f).
Notably, $B_{c2}^{\|}$ significantly exceeds the Pauli paramagnetic limit ($B_{p}$ = 1.86 $T_{c}$), reaching $B_{c2}^{\|}$/$B_{p}$ $\sim$ 3.56 at 1.15 K for the bulk sample S15.
This exceptionally large $B_{c2}^{\|}$ suggests the presence of Ising SOC or SOS.

\begin{figure*}[!thb]
\begin{center}
\includegraphics[width=6.5in]{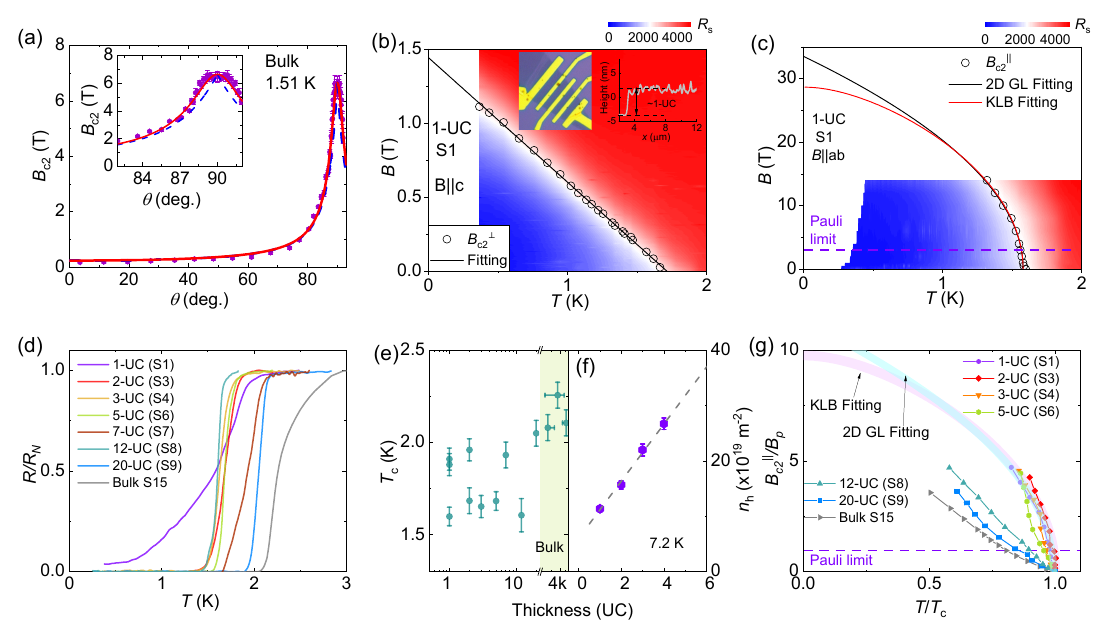}
\end{center}
\caption{\label{Fig2} Thickness-dependent superconductivity.
(a) Angular dependent $B_{c2}$ at 1.51 K for the 3-UC device. The data above 90$^{\circ}$ is mirrored from the data below 90$^{\circ}$ for clarity of presentation.
(b) $R_s$ map as a function of $T$ and $B^{\perp}$ for the 1 UC device S1, showing the extracted $B_{c2}^{\perp}$ together with the corresponding 2D GL fit.
The inset shows an optical photograph of the device and its thickness measurement.
(c) $T$-dependent $R_s$ measured under varying $B^{\|}$.
The extracted $B_{c2}^{\|}$ are fitted by 2D GL and KLB model.
(d) $T$-$R/R_{N}$ curves for devices of varying thickness.
(e) Thickness-dependent $T_{c}$.
(f) $n_h$ as a function of thickness, with the dashed guide line.
(g) $T/T_{c}$-dependent $B_{c2}^{\|}/B_{p}$ at different samples.
The representative KLB and 2D GL fitting results for 1-UC S1 are shown. }
\end{figure*}

Bulk band-structure calculations [Fig. 1(g)] support the presence of Ising SOC.
The broken inversion symmetry together with strong SOC induces spin polarization and splitting near the K point, giving rise to Ising-type SOC that has been reported in monolayer systems  \cite{MoS2IsingSC_Science15,MoS2IsingSC_NP16,NbSe22D_XiX_NP16}.
The band structure of a single unit cell of InBr(TaSe$_{2}$)$_{2}$ closely resembles that of the bulk, likely owing to weak interlayer coupling.
Similar spin-split features are also observed in both bulk and bilayer states of 2Hb-TaSe$_2$ \cite{TMDStackingPhase_PRB2004} (Fig. S3).
These results suggest that the 2Hb-TaSe$_2$ bilayer unit may help stabilize the spin-splitting feature in the bulk regime.
Furthermore, we synthesized InBr-intercalated monolayer TaSe$_2$, and found that a substantially enhanced $B_{c2}^{\|}/B_{p}$ emerges only in the bilayer bulk system InBr(TaSe$_2$)$_2$ (See Supplementary Note 8).
This indicates that, even in heterostructures containing similar interlayer spacers and 2D superconducting layers, achieving a large $B_{c2}^{\|}/B_{p}$ in the bulk regime remains not easy.
The 2Hb-TaSe$_2$ bilayer motif may offer a structural clue for identifying superconductors with a large $B_{c2}^{\|}/B_{p}$, though further studies are needed to confirm its broader applicability.

Anisotropic $B_{c2}$ behavior is measured in Fig. 2(a) for bulk InBr(TaSe$_{2}$)$_{2}$. The angular dependence $B_{c2}$($\theta$) is fitted using both the 3D anisotropic Ginzburg-Landau (AGL) model and the 2D Tinkham model (see Supplementary Note 7). The inset of Fig. 2(a) shows a better agreement with the 3D AGL model, exhibiting characteristics typical of conventional 3D superconducting nature for its bulk state.

\vspace{3ex}
\noindent\textbf{2.2 Thickness-dependent superconductivity}

Figure 2 explores the thickness-dependent superconductivity in InBr(TaSe$_{2}$)$_{2}$ to further study the exceptionally large $B_{c2}^{\|}$ behavior.
Temperature-dependent sheet resistance ($R_s$) mappings under out-of-plane ($B^{\perp}$) and in-plane ($B^{\|}$) magnetic fields for the 1-unit-cell (UC) device S1 (thickness given in unit cells, 1 UC = $c$ = 57.353 \AA) are presented in Figs. 2(b) and 2(c), respectively. The extracted $B_{c2}$ behavior can be well fitted by the Klemm-Luther-Beasley (KLB) model \cite{SOScatterSC_PRB1975} or the 2D Ginzburg-Landau (GL) model \cite{MoS2IsingSC_NP16}.
With increasing thickness, the $R$-$T$ curves [Fig. 2(d)] reveal a gradual increase of $T_{c}$ [Fig. 2(e)], consistent with the behavior typical of conventional 2D superconductors \cite{2DSC_SatioY_NRM2017,NbSe22D_XiX_NP16,2HNbSe2TaSe2SC_NC2018,1TdMoTe2SC_NC2019,MoS2GatedSC_NN2016,3RTaSe2SC_NanoLett2020}.
For thicknesses below 5 UC, the $T/T_{c}$-dependent $B_{c2}/B_{p}$ [Fig. 2(g)] can be also described by the KLB \cite{SOScatterSC_PRB1975} or 2D GL model \cite{MoS2IsingSC_NP16} (see Figs. S4 and S5).
Notably, the $B_{c2}^{\|}/B_{p}$ ratio exceeds 4 and changes slightly below 5 UC ($\sim$ 29 nm) [Fig. 2(g)], in stark contrast to conventional Ising or non-Ising superconductors \cite{NbSe22D_XiX_NP16,2HNbSe2TaSe2SC_NC2018,1TdMoTe2SC_NC2019,3RTaSe2SC_NanoLett2020,ThickGS_HuangCe_SB21},
in which $B_{c2}^{\|}/B_{p}$ usually declines rapidly with increasing thickness \cite{FFLOSC_JSPS07}.

\begin{figure*}[!thb]
\begin{center}
\includegraphics[width=6.5in]{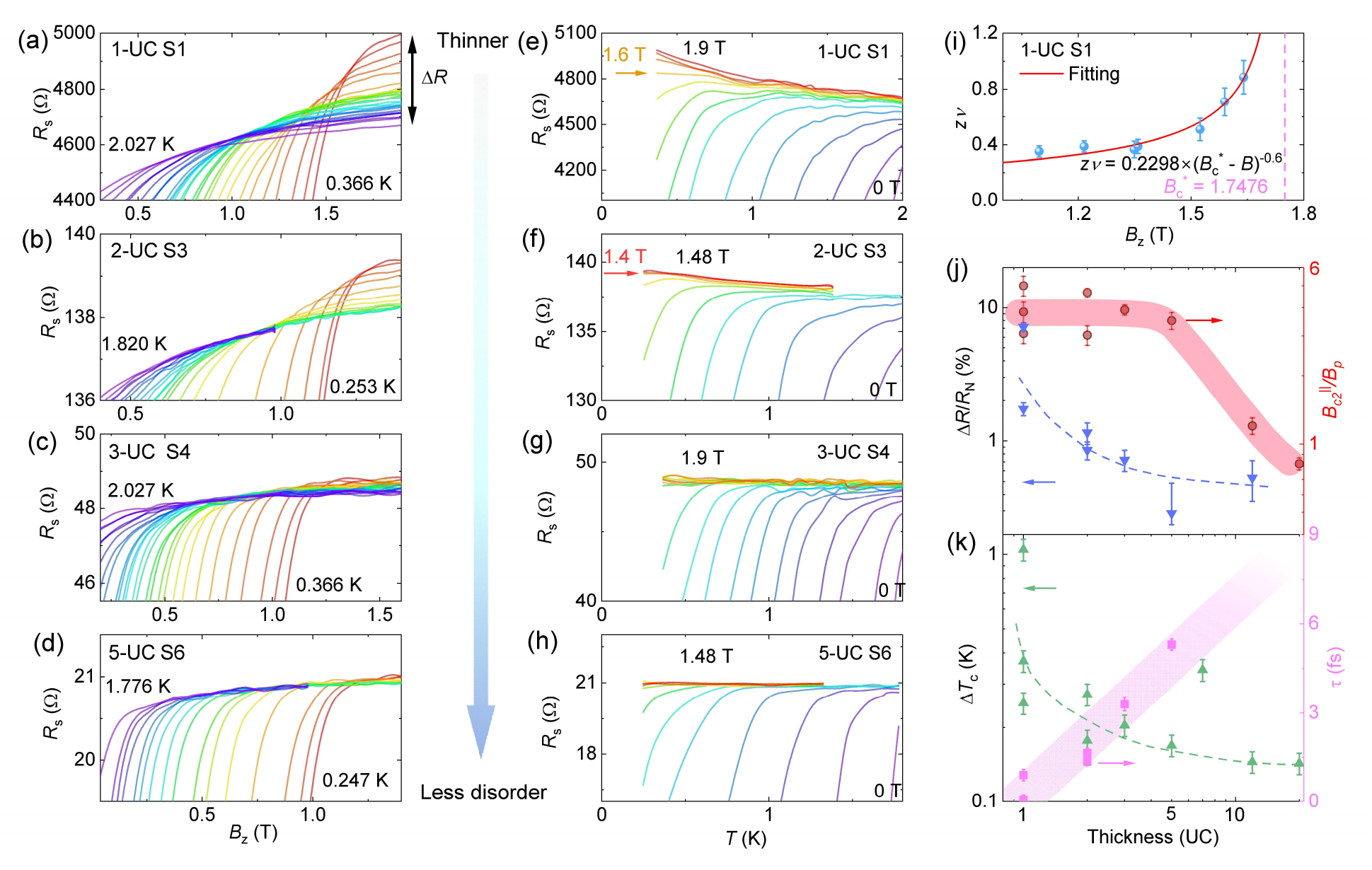}
\end{center}
\caption{\label{Fig4} $R_{s}$-$B_z$ (a-d) and $R_{s}$-$T$ curves (e-h) for different InBr(TaSe$_{2}$)$_{2}$ flakes. (i) $B$-dependent $zv$ with theoretical fits for 1-UC flakes. The error bar of $zv$ is obtained during the scaling analysis.
(j,k) Thickness-dependent resistance enhancement ratio $\Delta R/R_{N}$ (blue scatters, blue dashed guide line), $B_{c2}^{\|}/B_{p}$ [red scatters extracted from Fig. 2(g) at $T/T_c$ = 0.88, red guide line], superconducting transition width $\Delta T_{c}$ = $T_{c}^{90\%}$ - $T_{c}^{10\%}$ (green scatters, green dashed guide line), $\tau$ (magenta guide range). }
\end{figure*}

The further superconductivity is analyzed below. Hall measurements [Fig. 2(e) and Fig. S2] reveal a typical linear correlation between hole carrier density ($n_h$) and film thickness \cite{ThickvsN_APL2014}, allowing estimation of several physical parameters.
For instance, in the 1-UC flake S2, the SOS time $\tau_{so}$ obtained from KLB fitting \cite{MoS2IsingSC_Science15} is about 29.7 fs, while the total scattering time $\tau$ is significantly shorter ($\sim$ 0.9 fs). This indicates that the system operates in the dirty-limit regime.
Furthermore, finite-momentum pairing can be ruled out as a dominant mechanism for the large $B_{c2}^{\|}$, as the enhancement it provides is typically capped at $\sqrt{2}B_{p}$ \cite{FFLOSC_JSPS07}.

\begin{figure*}[!thb]
\begin{center}
\includegraphics[width=6.5in]{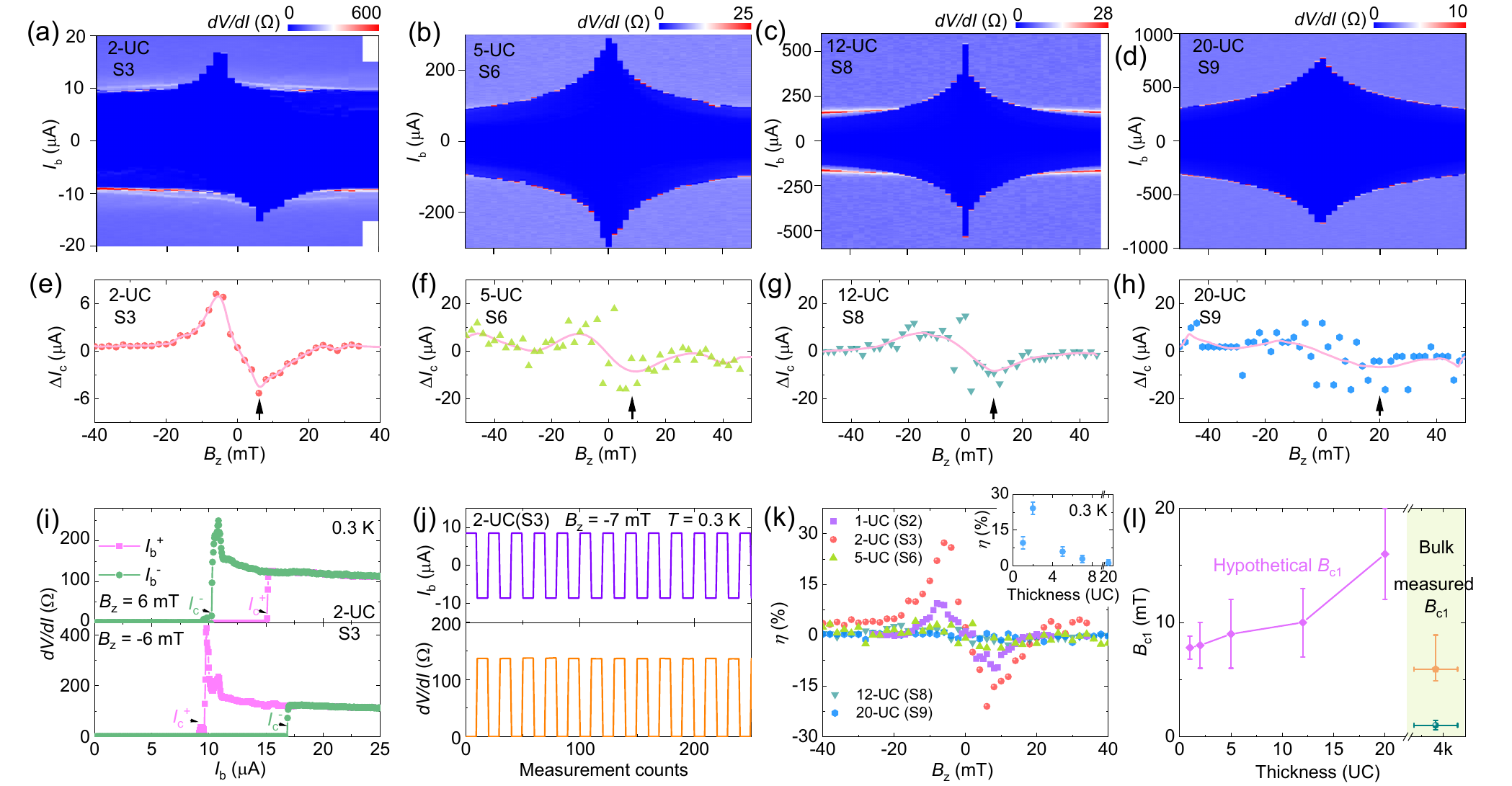}
\end{center}
\caption{\label{Fig4} (a-d) $dV/dI$ map as a function of $I_b$ and $B_z$ for 2, 5, 12 and 20 UC at 0.3 K.
(e-h) $B_z$-dependent $\Delta I_{c}$. A smooth guide curve (pink line) is included to clarify the overall trend.
(i) $dV/dI$-$I_{b}$ curves for positive and negative current sweep under $\pm$6 mT.
(j) Superconducting half-wave rectification at $B_{z}$ = -7 mT and 0.3 K in the device S3.
(k) $B_z$-$\eta$ behaviors at different thicknesses.
The inset displays the thickness-dependent $\eta$, with values taken as the maximum magnitude in panel (k). (l) Thickness-dependent $B_{c1}$. Magenta points are extracted from the peak positions indicated by black arrows in panels (e-h). Cyan and orange symbols represent $B_{c1}$ values obtained under $B^{\|}$ and $B^{\perp}$ in Bulk S15 (see Fig. S2), respectively.
    }
\label{Fig4}
\end{figure*}

\vspace{3ex}
\noindent\textbf{2.3 Superconductor-metal transition}

This section further examines the role of disorder effect/SOS \cite{SOScatterSC_PRB1975}, which is especially pronounced in dirty 2D superconductors where substantial disorder leads to $\tau$ is shorter than $\tau_{so}$. In such systems, quenched disorder typically accompanies superconductor-to-metal transitions \cite{GriffithsGafilm_science15}, thereby offering a pathway to study the thickness evolution of disorder and SOS \cite{ThickGS_HuangCe_SB21}.
Figures 3(a-d) show the $R_s$-$B_z$ curves for flakes ranging from 1 to 5 UC in thickness. Multiple crossing points in the curves indicates a transition from superconductivity to a weakly localized metallic state, as further illustrated in Figs. 3(e-h). Moreover, Figs. 3(i) and Fig. S8 reveal that the critical exponent $zv$ follows an activated scaling behavior, $zv \propto |B_{c}^{*}-B|^{-0.6}$, suggesting the presence of a Griffiths singularity \cite{GriffithsGafilm_science15} in flakes thinner than 2 UC. In contrast, 3-UC (S4) and 5-UC (S6) flakes exhibit conventional superconducting transitions [Figs. 3(c,d,g,h) and Figs. S6-S8].
At low temperatures, magnetic-field-induced resistance enhancement can be roughly quantified as $\Delta R/R_{N}$ = [$R$ ($\sim$ 0.3 K) - $R$ ($\sim$ 2 K)]/$R$ ($\sim$ 2 K)], shown in Fig. 3(j). The $\Delta R/R_{N}$ decreases with thicker films, implying that disorder and SOS effects become increasingly prominent in thinner flakes.

This trend is also supported by the broadening of the superconducting transition $\Delta T_c$ (green scatters) with decreasing thickness, as shown in Fig. 3(k). Further estimation of the disorder-related scattering time $\tau$ shows the smaller value in thinner films,
indicating stronger disorder effect and enhanced SOS \cite{ThickGS_HuangCe_SB21}.
The enhanced SOS effect should contribute to the larger $B_{c2}^{\|}$, considering the same interlayer coupling strength in these films \cite{SOScatterSC_PRB1975}.
Nevertheless, the nearly monotonic enhancement of the disorder contrasts with the approximately invariant $B_{c2}^{\|}/B_p$ below 5 UC [red scatters in Fig. 3(j), and Fig. 2(f)]. This disparity suggests that the exceptionally large $B_{c2}^{\|}/B_p$ - especially in flakes below 5 UC - is primarily driven by 3D Ising SOC rather than by SOS.
However, quantifying the respective contributions of these two mechanisms to the enhanced $B_{c2}^{\|}$ remains an open challenge in this class of materials and calls for further investigation.

\vspace{3ex}
\noindent\textbf{2.4 Superconducting diode effect}

The SDE is further measured to probe Ising SOC in InBr(TaSe$_{2}$)$_{2}$, because SOS cannot give rise to SDE.
The $dV/dI$ map as a function of bias current $I_b$ and $B_z$ are shown in Figs. 4(a-d) for different thicknesses. The antisymmetric behavior of $\Delta I_c = I_{c+} - I_{c-}$ [Figs. 4(e-h)] confirms the presence of SDE, where critical current $I_{c+}$ and $I_{c-}$ is obtained from sweeping positive and negative $I_b$ in Figs. 4(a-d), respectively. Figure 4(i) further displays the antisymmetric SDE in 2-UC (S3) flake at $\pm$6 mT, and stable superconducting half-wave rectification is observed under an alternating $I_{b}$ = $\pm$8.5 $\mu$A [Fig. 4(j)].
A similar SDE is also observed for the 1-UC flake (Fig. S9).
With increasing thickness, the antisymmetry in $\Delta I_c$ gradually weakens [Figs. 4(e-h)],
and a smooth guide line (pink line) is added to highlight this trend.

The SDE efficiency is quantified by the ratio $\eta = (I_{c+} - |I_{c-}|)/(I_{c+} + |I_{c-}|)$ \cite{JDETa2Pd3Te5_NC2024} for comparison across different films [Fig. 4(k)].
The apparent SDE is observed in flakes thinner than 12 UC ($\sim$ 69 nm), which can be attributable to Ising SOC \cite{NbSe2SCE_BauriedlL_NC2022}, as discussed in detail below.
Moreover, the persistence of antisymmetric $\Delta I_c$ between 5-UC and 12-UC flakes indicates that IS remains effective in bulk-like range.
However, for thicknesses exceeding 20 UC, SDE nearly vanishes [Fig. 4(h)],
despite bulk band calculations predicting substantial spin polarization [Fig. 1(g)].
The suppression of the SDE in thicker films may be attributed to slightly weakening of the spin splitting or a moderate increase in defects like stacking faults, a common feature in intercalated TMDs \cite{TMDStackingPhase_PRB2004}.
Such effects may strengthen orbital limiting effects, consequently reducing the $B_{c2}^{\|}/B_{p}$ ratio above 5 UC [Fig. 2(d)]. Notably, the ratio remains above 3 even in bulk, indicating that the defects or weakening of the spin splitting are not severe.

Furthermore, other typical mechanisms that can produce antisymmetric SDE are analyzed and ruled out: (1) Superconducting interference mechanism is first excluded in this intrinsic superconductor \cite{JDETa2Pd3Te5_NC2024}; (2) The material exhibits no signature of magnetism \cite{NbVCoVTaFFSDE_NN2022} or chiral domains \cite{CsVSbSDE_Nature2024}; (3) Mechanisms relying on Rashba SOC or finite-momentum Cooper pairing \cite{NbTaVSDE_AndoF_nature2020,NiTe2JDE_PalB_NP2022} are also excluded, as they typically require an in-plane magnetic field, which is not the case here; (4) The vortex magnetism generally needs relatively large magnetic field to break lower critical field ($B_{c1}$). Below the $B_{c1}$, the contribution of the Meissner screening effect to the SDE originating from asymmetric edges of the flake \cite{VFilm_HouYS_arXiv2022} is analyzed and ruled out in our flakes.
If the SDE here were solely attributable to this mechanism,
the peak positions marked by black arrows in Figs. 4(e-h) correspond to a lower critical field ($B_{c1}$) \cite{VFilm_HouYS_arXiv2022}.
Above this field, vortex penetration is expected to commence, leading to the suppression of the SDE.
In Fig. 4(l), the thickness dependence of this hypothetical $B_{c1}$ is summarized by magenta scatters. It shows a gradual increase as thickness increases, which exhibits a trend opposite to the characteristic rapid decrease of $B_{c1}$ in thin-film superconductors \cite{Hc1withThickness_APLM15,Bc1vsFilm_PRB1994}.
Furthermore, the experimentally determined bulk $B_{c1}$ values (cyan and orange scatters, obtained under $B^{\|}$ and $B^{\perp}$ in Fig. S2) deviate from the extrapolated trend of the hypothetical $B_{c1}$.
Additionally, SDE induced by the Meissner screening effect seems to diminish
and vanishes in thicker films \cite{SDErectifier_NE15}.
For instance, no evident antisymmetric $\Delta I_c$ is observed in Nb films ($\sim$ 60 nm thick) without specialized structural design (Fig. S10).
However, in our flakes, an antisymmetric $\Delta I_{c}$ persist beyond 12-UC ($\sim$ 69 nm), as shown in Fig. 4(g). In short, the Meissner screening effect is unlikely to be the primary origin of the SDE here, and we instead attribute it to Ising SOC.

\vspace{3ex}
\noindent\textbf{3 Conclusion}

In summary, we synthesized the bulk noncentrosymmetric superconductor InBr(TaSe$_2$)$_2$. A large $B_{c2}^{\|}/B_p$ is obtained in both bulk and thin-film forms, raising the central question of whether its origin stems from Ising SOC or SOS.
Through semi-quantitative analysis, we demonstrate that Ising SOC plays a significant role.
This conclusion is supported by multiple lines of evidence:
This conclusion is supported by multiple lines of evidence: (i) the bulk state exhibits 3D superconductivity and a large $B_{c2}^{\|}$; (ii) first-principles band structure calculations confirm the presence of Ising SOC in both bulk and thin-film limits; (iii) the thickness-dependent $B_{c2}^{\|}/B_p$ behavior deviates from the SOS mechanism; and (iv) the SDE is attributed to Ising SOC rather than to other known mechanisms.
Our findings also show the thickness-dependent characteristics of bulk Ising superconductivity.

\vspace{3ex}
\noindent\textbf{4 Experimental Section}

\noindent\textbf{4.1 Crystal Growth.}

The single crystals InBr(TaSe$_{2}$)$_{2}$ were prepared successively using the
solid-state reaction method and the vapor transport method, respectively,
referred to In$_{x}$TaSe$_{2}$ \cite{InTaSe2_YpLi}.
Firstly, the stoichiometric mixture of In (Alfa Aesar 99.95\%), Ta (Alfa Aesar 99.95\%)
and Se (Alfa Aesar 99.95\%) was sealed in a quartz ampoule and then heated to
1123 K for 2 days. Secondly, the resultant was reground with InBr$_{3}$ and then sealed in an evacuated ampoule with a length of 16 cm. The InBr$_{3}$ here was used as a transport agent ($\sim$ 6 mg/cm$^{3}$). Several platelike single crystals were grown in a two-zone furnace for
three weeks with the powder end at 1143 K and the cooler end at 1023 K.
InBrTaSe$_{2}$ single crystals were grown analogously, with a slight increase in synthesis temperature.

\vspace{3ex}

\noindent\textbf{4.2 Device fabrication.}

InBr(TaSe$_{2}$)$_{2}$ thin films were all obtained by mechanically exfoliating bulk sample S15 onto SiO$_{2}$/Si substrates, followed by coating with PMMA at 4000 rpm for 60 seconds in a glove box and subsequent annealing at 120 $^{\circ}$C for 180 seconds.
Electrical contacts were patterned using electron-beam lithography and subsequent development.
Ti/Au electrodes were deposited after Ar ion etching was performed to remove the surface oxide layer. Notably, the Ar ion etching step is essential for achieving effective electrode contact in this material system. After the lift-off process, the device was encapsulated with a PMMA layer to prevent oxidation. The entire fabrication procedure was carried out in a nitrogen atmosphere glove box, following a methodology similar to that reported previously \cite{Ta2Pd3Te5LL_NC2023}.
The film thicknesses were measured using atomic force microscope.

\vspace{3ex}

\noindent\textbf{4.3 Transport measurements.}

Electrical transport measurements were conducted using cryostats (Oxford Instruments refrigerator). DC current bias was applied via a Keithley 2612 current source. The resistance of thin film devices was determined using a standard low-frequency (7-14 Hz) lock-in amplifier (LI5640, NF Corporation), whereas the resistance of bulk samples was measured using a DC technique employing a Keithley 2400 source meter in combination with a Keithley 2182 nanovoltmeter. 

\vspace{3ex}

\noindent\textbf{4.4 Structure characterization.}

The structure was characterized by SXRD and HRTEM. SXRD measurements were carried out on the a Bruker D8 Venture diffractometer with a Mo-K$_\alpha$ radiation. The structure was solved by a direct method and further refined by full-matrix least-squares algorithm of $F^2$ using the software package SHELXTL \cite{SHELX}. HRTEM measurements were performed at room temperature using an aberration-corrected FEI-Titan G2 80-200 ChemiSTEM. The compositional ratios of the samples were analyzed using EDS and ICP-AES. Based on the combined data presented in Supplementary Note 1, the chemical composition of the superlattice is proposed to be In:Br:Ta:Se in a 1:1:2:4 ratio.

\vspace{3ex}

\noindent\textbf{4.5 Band structure calculations.}

We performed density functional theory (DFT) calculations using the Vienna ab initio simulation package (VASP) \cite{VASP1,VASP2} in conjunction with the projector augmented wave (PAW) method \cite{PAW1994,PAW1999}. The exchange-correlation effects were treated within the generalized gradient approximation (GGA) using the Perdew-Burke-Ernzerhof (PBE) functional \cite{PBE}. The cut-off energy of plane wave expansion was set to 350 eV. The Brillouin zone was sampled by $12\times12\times1$ $k$-points mesh in the self-consistent process. The spin polarization $S_z$ was computed using VASP and visualized with PyProcar \cite{pyprocar1,pyprocar2}.

\vspace{3ex}

\noindent\textbf{Acknowledgments}

We thank Youting Song for the assistance with the single-crystal X-ray diffraction measurements. This work was supported by the National Natural Science Foundation of China (Grant Nos. 12404154, 92065203, 12174430, 12174334, 12204298, 12304080), the Beijing Natural Science Foundation (Grant No. JQ23022), the HZNU scientific research and innovation team project (No. TD2025013), the Strategic Priority Research Program of Chinese Academy of Sciences (Grant No. XDB33000000), the CAS Superconducting Research Project (Grant No. SCZX-0101), Beijing National Laboratory for Condensed Matter Physics (Grant Nos. 2023BNLCMPKF019 and 2025BNLCMPKF013), the National Key Research and Development Program of China (Grant No. 2022YFA1403202), the Hangzhou Joint Fund of the Zhejiang Provincial Natural Science Foundation of China (Grant No. LHZSZ24A040001), and the Zhejiang Provincial Natural Science Foundation of China (Grant No. LMS26A040010).
A portion of this work was carried out at the Synergetic Extreme Condition User Facility (SECUF).

%
%

\vspace{3ex}

\noindent\textbf{Conflicts of Interest}

\noindent The authors declare no competing interests.

\vspace{3ex}

\noindent\textbf{Data Availability Statement}

\noindent The data that support the findings of this study are available
from the corresponding author upon reasonable request.

\vspace{3ex}

\end{document}